\documentclass[a4paper, 10pt, conference]{IEEEtran}
\IEEEoverridecommandlockouts
\usepackage{cite}
\usepackage{amsmath,amssymb,amsfonts}
\usepackage{algorithmic}
\usepackage{graphicx}
\usepackage{textcomp}
\usepackage{xcolor}
\def\BibTeX{{\rm B\kern-.05em{\sc i\kern-.025em b}\kern-.08em
    T\kern-.1667em\lower.7ex\hbox{E}\kern-.125emX}}
\begin{document}
\bstctlcite{IEEEexample:BSTcontrol}
\title{Thailand Asset Value Estimation Using Aerial or Satellite Imagery
}

\author
{\IEEEauthorblockN{Supawich Puengdang}
\IEEEauthorblockA{\textit{KASIKORN Labs} \\
\textit{KASIKORN Business-Technology Group}\\
Nonthaburi, Thailand \\
supawich@gmail.com}
\and
\IEEEauthorblockN{Worawate Ausawalaithong}
\IEEEauthorblockA{\textit{Department of Bio and Brain Engineering} \\
\textit{Korea Advanced Institute of Science and Technology}\\
Daejeon, South Korea \\
a.worawate@kaist.ac.kr}
\and
\IEEEauthorblockN{Phiratath Nopratanawong}
\IEEEauthorblockA{\textit{KASIKORN Labs} \\
\textit{KASIKORN Business-Technology Group}\\
Nonthaburi, Thailand \\
phiratath.n@kbtg.tech}
\and
\IEEEauthorblockN{Narongdech Keeratipranon}
\IEEEauthorblockA{\textit{Department of Computer Engineering} \\
\textit{Chulalongkorn University}\\
Bangkok, Thailand \\
narongdech.k@chula.ac.th}
\and
\IEEEauthorblockN{Chayut Wongkamthong}
\IEEEauthorblockA{\textit{KASIKORN Labs} \\
\textit{KASIKORN Business-Technology Group}\\
Nonthaburi, Thailand \\
chayut.w@kbtg.tech}
}

\maketitle

\begin{abstract}
Real estate is a critical sector in Thailand's economy, which has led to a growing demand for a more accurate land price prediction approach. Traditional methods of land price prediction, such as the weighted quality score (WQS), are limited due to their reliance on subjective criteria and their lack of consideration for spatial variables. In this study, we utilize aerial or satellite imageries from Google Map API to enhance land price prediction models from the dataset provided by Kasikorn Business-Technology Group (KBTG). We propose a similarity-based asset valuation model that uses a Siamese-inspired Neural Network with pretrained EfficientNet architecture to assess the similarity between pairs of lands. By ensembling deep learning and tree-based models, we achieve an area under the ROC curve (AUC) of approximately 0.81, outperforming the baseline model that used only tabular data. The appraisal prices of nearby lands with similarity scores higher than a predefined threshold were used for weighted averaging to predict the reasonable price of the land in question. At 20\% mean absolute percentage error (MAPE), we improve the recall from 59.26\% to 69.55\%, indicating a more accurate and reliable approach to predicting land prices. Our model, which is empowered by a more comprehensive view of land use and environmental factors from aerial or satellite imageries, provides a more precise, data-driven, and adaptive approach for land valuation in Thailand.
\end{abstract}

\begin{IEEEkeywords}
Land price prediction, Aerial and satellite imagery, Computer vision, Siamese Neural Network, Weighted quality score
\end{IEEEkeywords}

\section{Introduction}
Real estate is a critical sector in the economy of any country, and Thailand is no exception. The growth of this sector has led to a surge in the demand for accurate land price prediction models. Not only is an accurate pricing method important to real estate but it would also benefit other fields such as finance and urban planning. It provides stakeholders with valuable information to make informed decisions about the value of a piece of land and the opportunity for investment \cite{realestate}.

\indent
Traditional methods of land price prediction in Thailand typically rely on weighted quality scores (WQS) \cite{wqs}, which are assigned by appraisers. WQS is based on various factors such as the location, distance from key amenities, prices of nearby lands, and the condition of the building on the land. These factors are scored and then combined to get the overall score for each piece of land. However, this method can be limited due to, firstly, its reliance on subjective criteria with possible biases from appraisers. Secondly, it focuses only on tabular data consisting of historical sales, location, land use, and zoning regulations and lacks consideration for spatial variables such as changes in geography, infrastructure, and environmental factors.

\indent
Machine learning and deep learning models can learn patterns and relationships from large amounts of data and identify important features that are used in the predictions. This leads to its breakthrough in many use cases including computer vision, semantic analysis, natural language processing, etc. \cite{ml}. There have been attempts to use machine learning to perform land price prediction; however, with just tabular data, the results are still not convincing \cite{london, la, japan}.

\indent
With the growth of deep learning, image-based applications have become more accurate and effective, leading to breakthroughs in fields such as image classification, object detection, semantic segmentation, etc. \cite{cnn}. The high dimensionality of image data and its ability to capture complex patterns and structures make it an ideal input for many machine learning tasks. By training on large sets of labeled images, machine learning models can learn to recognize and classify objects, detect patterns, and extract necessary information. 

\indent
Recent research has explored the potential of combining satellite imagery and tabular data to predict land prices using convolutional neural networks (CNNs). CNNs are a type of deep learning algorithm commonly used in image recognition tasks, which can learn complex features and patterns from images. In this approach, satellite imagery is used as an additional input to the traditional tabular data in predicting land prices. In addition, satellite imagery is preprocessed and transformed into numerical data using image processing techniques, such as image segmentation and feature extraction. These additional features can help leverage the overall performance of the model. A number of studies have shown that, by including satellite imagery in the prediction, the prediction result improves substantially\cite{london, la, japan}. However, in Thailand, the use of such imagery in land valuation is still limited and unexplored.

\indent
Therefore, in this study, we aim to explore the utilization of aerial or satellite imagery in performing land price prediction in Thailand and compare its performance with traditional machine learning methods that do not use the images. For land price prediction, aerial or satellite imagery may provide a more comprehensive view of land use and environmental factors. The accessibility of such imagery and advancements in machine learning algorithms allow for the integration of various spatial data sets, such as accurate land boundaries, landmarks, and surrounding infrastructures, into land value prediction models. With the help of imagery, land value prediction in Thailand can become more precise, data-driven, and adaptive to changing conditions. 

\indent
The remainder of this article is organized as follows. In section 2, we explore related works, such as similarity learning and EfficientNet. The subsequent sections provide a more detailed discussion of our methodology, including the data used in this work, preprocessing techniques that we utilized, traditional machine learning methods, deep learning methods, and the ensemble of the two models. In section 5, we present the results and discussion. Lastly, the conclusion of our work is in section 6. From now on, for conciseness, we refer to aerial or satellite imagery as satellite imagery.

\section{Related Works}

\subsection{House Price Estimation}

There have been works that utilize satellite imagery in house price prediction. The work in \cite{london} uses CNN to model geospatial data of house prices in London, Birmingham, and Liverpool. They use CNNs to extract features from satellite images of regions around the test sample. These features are then combined with the tabular data (house attributes) and fed to regressors to predict the house price.  The work in \cite{la} incorporates satellite imageries with tabular data to predict house prices in Los Angeles using an Inception-v3 model pretrained on ImageNet. They combined the feature vectors of the image and the tabular data at the last layer of the network. Satellite imageries have shown a significant impact of 10\% improvement in the R-squared score in this work.

\subsection{Land Price Estimation}

The work in \cite{japan} created three models to compare the effectiveness of satellite images on land price prediction: only tabular data, only satellite images, and both combined. The combined model achieved the best accuracy from 2 out of 3 Japanese land prices datasets.

\subsection{Siamese Network/Similarity tasks}
Similarity learning is a popular area of research in machine learning, where the goal is to learn a similarity metric between objects in a given space. One of the common approaches to similarity learning is the Siamese network \cite{siamese}, which is a type of neural network architecture that takes a pair of inputs and learns to output a similarity score between them. The Siamese network consists of two identical subnetworks that share the same weights and are fed with the input pairs. By minimizing a contrastive loss function, the Siamese network learns to map similar input pairs close to each other and dissimilar ones far apart in the learned embedding space. Siamese networks have been used for various applications, such as image matching, face recognition, and sentence similarity, and have achieved state-of-the-art performance in many tasks.

\subsection{EfficientNet}
EfficientNet \cite{effnet} is a family of convolutional neural network architectures that were introduced in 2019. The key idea behind EfficientNet is to scale up neural networks in a more principled way, by balancing network depth, width, and resolution in a systematic manner. Specifically, EfficientNet uses a compound scaling method that uniformly scales the network depth, width, and resolution with a set of fixed scaling coefficients. By doing so, EfficientNet achieves state-of-the-art performance on various image classification benchmarks, while using significantly fewer parameters than the previous state-of-the-art models. Furthermore, EfficientNet has been shown to generalize well to other computer vision tasks, such as object detection and semantic segmentation. As a result, EfficientNet has become a popular choice of architecture in recent computer vision research.

\section{Methodology}
This section provides an overview of the methodology employed in our study. It begins with a discussion of the datasets used, followed by an explanation of the data preparation process. Subsequently, we provide descriptions of the deep learning model, machine learning model, and ensemble method implemented in our study.

\subsection{Data Sources}
The datasets utilized in our study consist of two types of data: tabular data and image data. The tabular data, obtained from the asset valuation information provided by land appraisers, includes a wide range of quantitative information related to the properties being evaluated, such as property characteristics, location details, facilities around the land, proximity to roads and alleys, and distance to the main street. Additionally, this data also includes the appraisal price of each piece of land, evaluated with the exclusion of the building on the land, and expressed in the unit of Thai Baht per square Wa with 1 square Wa equals 4 square meters. The tabular data was provided by Kasikorn Business Technology Group (KBTG) in a structured format and served as a valuable source of quantitative information for our analysis.

\begin{figure}[htbp]
\centerline{\includegraphics[width=0.9\columnwidth]{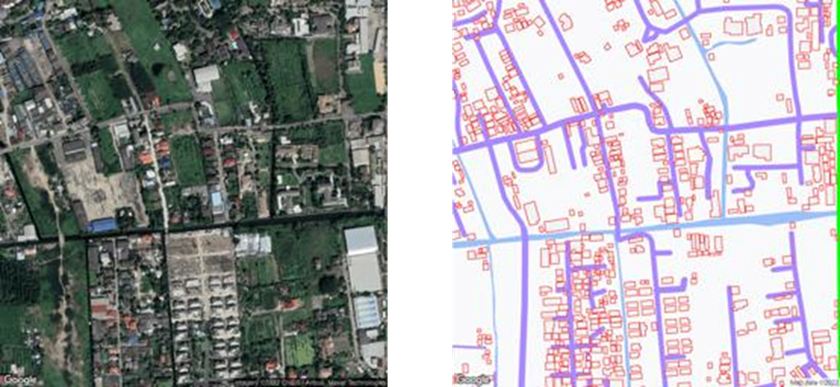}}
\caption{Satellite image (left) and Segmented image (right).}
\label{fig1}
\end{figure}

\indent
The image data used in our study was obtained from the Google Map API \cite{googlemap, googlemapAPI}, specifically satellite images and segmented images at the specific latitude and longitude coordinates of the land being appraised. The satellite images provided a visual representation of the land situated in the center of the image, along with its surrounding environment, covering an area of around $740\times740$ square meters. Meanwhile, the segmented images were preprocessed by Google \cite{googlemapAPI} to identify and label various features such as streets, alleys, buildings, water sources, and other relevant details, as shown in Fig.~\ref{fig1}. The segmented images were particularly useful in extracting visual features and contextual information for incorporation into our deep learning model to improve valuation accuracy.

\indent
In total, our dataset comprises approximately 11,000 data points, consisting of land data from 7 provinces located in the Northern, Central, Eastern, and Southern parts of Thailand. The dataset was further processed and analyzed in the subsequent sections of our study. The preprocessing steps involved cleaning, transforming, and integrating the tabular and image data to prepare it for the training and validation of our deep learning model.

\subsection{Data Preprocessing}
\begin{figure}[htbp]
\centerline{\includegraphics[width=1.0\columnwidth]{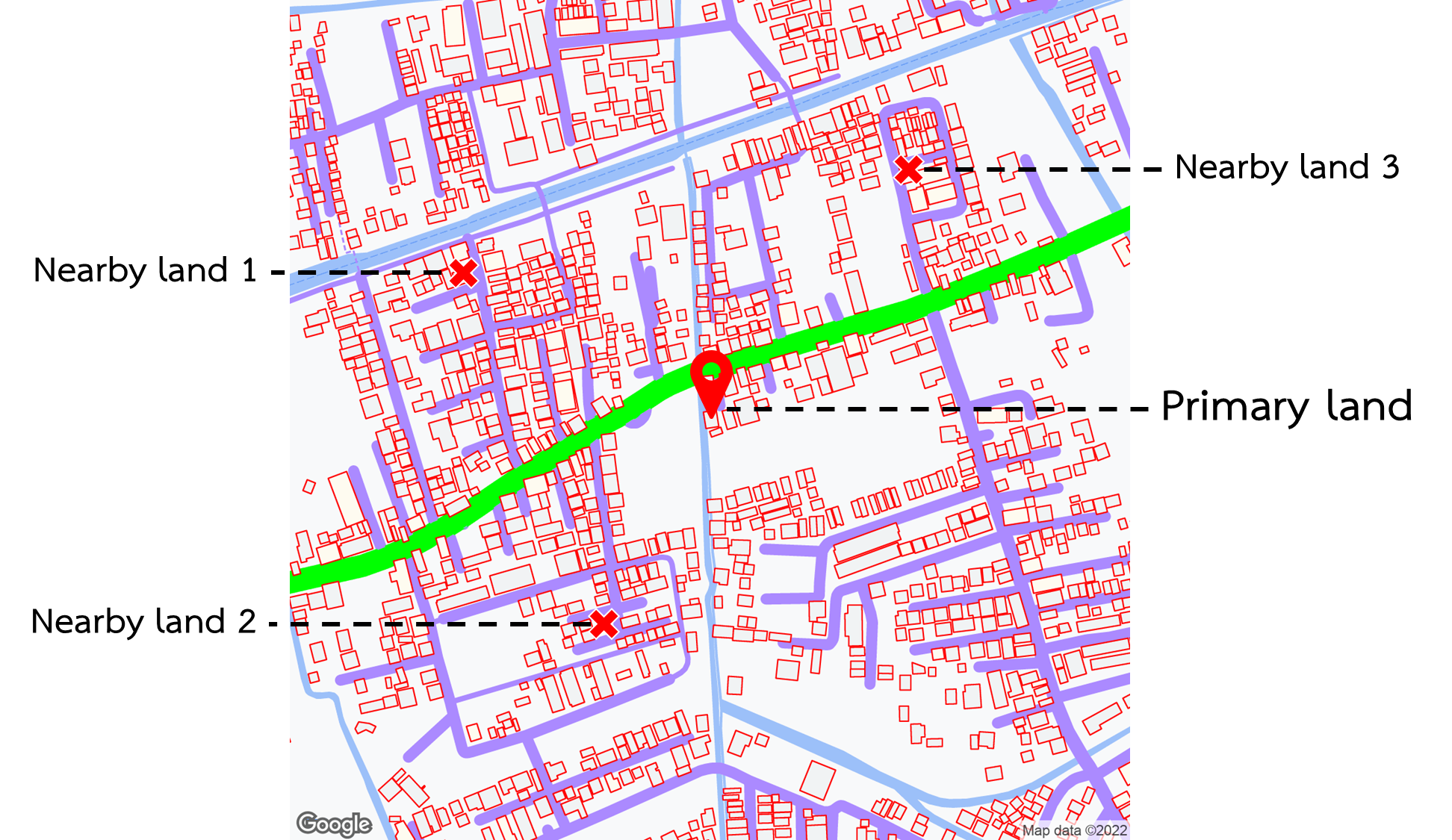}}
\caption{Illustration of nearby lands around the primary land.}
\label{fig2}
\end{figure}

We approach this problem as a classification task rather than a normal regression problem. As illustrated in Fig.~\ref{fig2}, land pairs that were close to each other within a radius of 3 kilometers were compared, and if the price difference did not exceed a certain threshold, they were labeled as the similar pair (1), and if it exceeded the threshold, they were labeled as the different pair (0), categorizing lands based on their similarity in terms of pricing.

\indent
For each pair of lands, each feature of the tabular data was compared. For continuous features, such as the distance to the main street and the area size of the land, the absolute difference between each feature was calculated. For categorical features, values were compared and encoded as the same or different. This allowed for capturing the variability in the features and their impact on land valuation.

\indent
In addition to the original features, additional features were manually created by comparing the pixel differences of various colors in the satellite imagery, such as greenness, blueness, and darkness. These color differences were hypothesized to be indicative of differences in land characteristics. These additional features aim to capture more nuanced information from the satellite images to improve the accuracy of the valuation model.

\indent
To refine the feature selection, basic tree-based models such as LightGBM \cite{lightgbm} and Random Forests \cite{rf} were used. All features were fed into the model, and feature importances were obtained for each feature. Based on the importance scores, only the most effective features were selected for use in the deep learning model, reducing the dimensionality of the dataset and focusing on the most informative features for the valuation task.

\indent
For the image data, additional preprocessing steps were performed. The images were resized into multiple square sizes, such as $128\times128$, $256\times256$, and $512\times512$ pixels, to ensure consistency in image resolution. Various image augmentation techniques, such as rotating, horizontal/vertical flipping, color jittering, and adding Gaussian Noise, were applied to augment the dataset and enhance model performance during experimentation. These techniques increased the diversity of the image data and improved the generalization capability of the deep learning model.

\indent
The datasets were split into a training set, a validation set, and a test set, following a ratio of 80-10-10, respectively, based on the primary land (the land set to find the price). The last 20\% of the dataset sorted by appraisal date from each province was reserved for use in the validation and test set randomly. The final dataset used for training our deep learning model consists of 96,028 data points, while the validation set comprises 23,178 data points, and the test set contains 26,783 data points. All datasets had a total of 286 features, which were used as input for our deep learning model.

\indent
This data-splitting approach ensured that we had a diverse and representative dataset for training, validation, and testing our model, while also accounting for regional differences in land appraisals. It helped us evaluate the performance and generalization of our model on unseen data during the validation and testing phases of our study.

\indent
Overall, the data preprocessing stage involves several steps, including the classification approach, feature comparison, manual feature creation, feature selection, and image data preprocessing. These steps ensure that the dataset is appropriately prepared and optimized for training the deep learning model for asset valuation.

\subsection{Machine Learning Approach}
To perform the classification task on our dataset, we employed several traditional machine learning methods, including logistic regression, k-nearest neighbors, support vector machines, decision-tree-based models, etc. In this approach, only the tabular data was used. To compare the impact of image features, we trained these models with and without our engineered features (such as pixel differences) manually created from the satellite images. In this work, the Extra Trees Classifier \cite{et} is used as the baseline as it performs the best with and without image data.

\subsection{Deep Learning Approach}
\begin{figure}[htbp]
\centerline{\includegraphics[width=0.9\columnwidth]{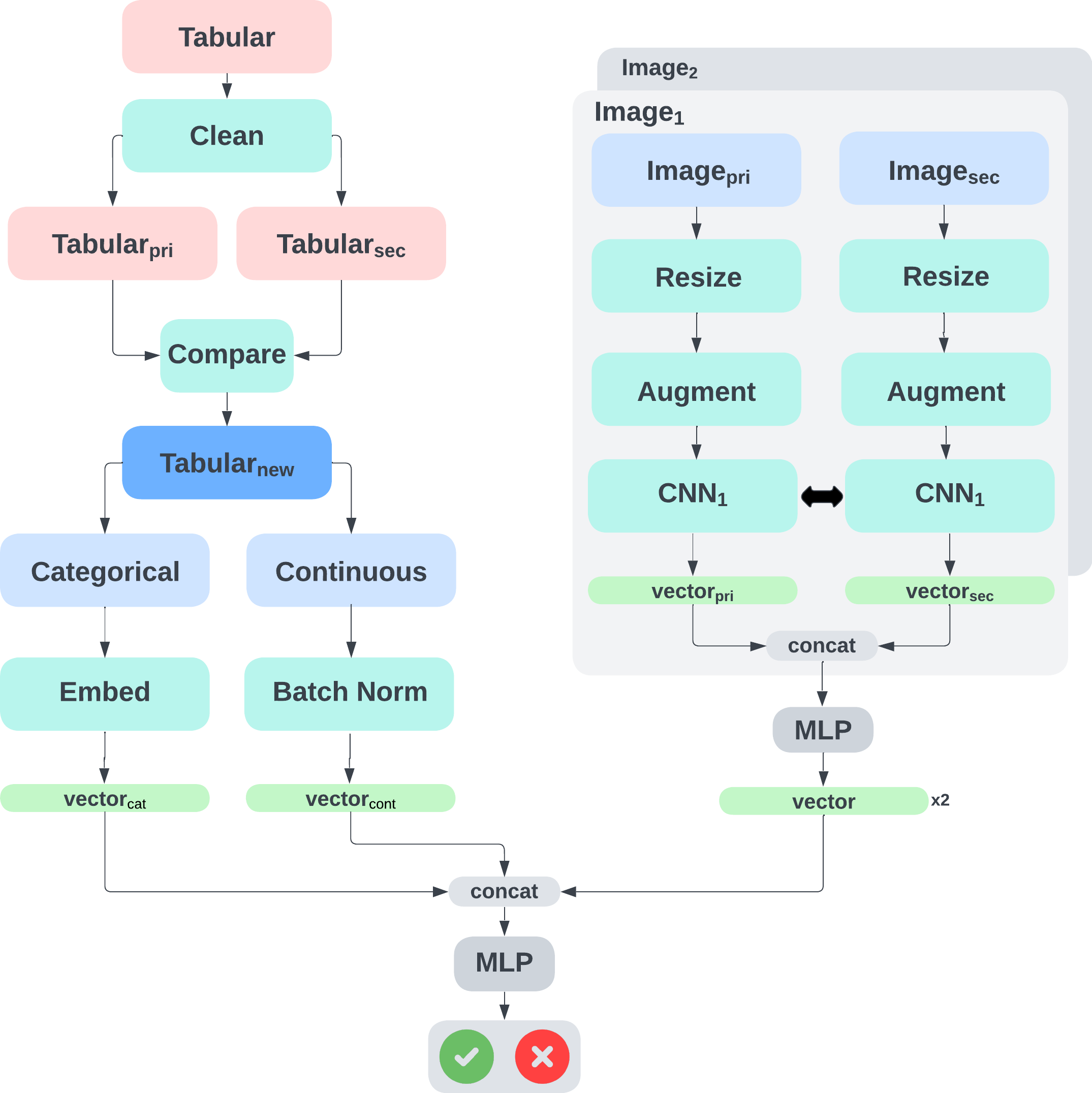}}
\caption{Illustration of the overall data preparation process and deep learning approach.}
\label{fig3}
\end{figure}

 Instead of using a straightforward method of an end-to-end regression model, we created a novel deep learning-based approach using a similarity model to assess the similarity score between pairs of lands, which was then used to determine the reasonable price of land based on nearby similar lands. Therefore, compared to the regression model, this technique brings more interpretability and transparency to the land valuation task. The entire pipeline is described in Fig.~\ref{fig3}.

\indent
Once the similarity scores were calculated for all the land pairs, they were compared with the predefined threshold. If the similarity score of a land pair was found to be higher than the threshold, it was considered similar enough, and the previous appraisal prices of those nearby similar lands were used for weighted averaging to predict the reasonable price of the land in question.

\indent
Our similarity learning scheme is inspired by the Siamese Neural Network. Each land pair --- primary and secondary land images, were fed to an identical image embedding CNN. A pretrained EfficientNet architecture was implemented as the embedding network. The newly created tabular data to compare two lands comprises of categorical features (province, street number, etc.) and continuous features (road width, color differences between image pairs, distances between pairs, etc.). The categorical features were embedded into a lookup table and later into a vector. Batch normalization \cite{batchnorm} was applied to the continuous features. To realize our custom multimodal model, we fed the concatenated embedded vectors into a multi-layer perceptron (MLP) to predict the final probabilities, which we consider as the similarity score for each pair.

\indent
During the training process, binary cross-entropy, a common loss function for binary classification tasks, was utilized as the loss function. The model was trained using the training dataset, and the model parameters were updated iteratively using backpropagation and gradient descent optimization.

\indent
To optimize the performance of the model, Optuna \cite{optuna}, a Python library for hyperparameter optimization, was employed. Various hyperparameters were experimented with, including the initial learning rate, a learning rate scheduler, an optimizer, dropout probability, and the number of hidden layers in the MLP classifier. Additionally, different pretrained EfficientNet sizes, ranging from b0 to b5, as well as EfficientNetV2 \cite{effnetv2} models of small to medium sizes, were evaluated. Furthermore, experiments were conducted to assess the impact of freezing the model layers on reducing the training time.

\indent
Overall, the deep learning-based approach with Siamese-inspired Neural Network allowed us to create a custom multimodal model for similarity-based asset valuation. The model was trained using binary cross-entropy loss and optimized using Optuna for hyperparameter tuning, providing a robust and accurate approach for predicting the reasonable price of lands based on their similarity to nearby lands.

\subsection{Ensemble Learning}
To enhance the similarity model, for this study, we implemented an ensemble technique that combined several machine learning models and deep learning models. The ensemble approach involved obtaining similarity scores from five different models, including two deep learning models and three tree-based models. The first deep learning model was trained using $128\times128$ pixels images, while the second deep learning model used $512\times512$ pixels images. Both of these models utilized EfficientNet-b1 as the image embedding architecture. The three tree-based models consisted of an Extra Trees classifier \cite{et}, a Random Forest classifier \cite{rf}, and another Random Forest classifier that was trained using the extracted latent space from the first deep learning model.

\begin{figure}[htbp]
\centerline{\includegraphics[width=0.9\columnwidth]{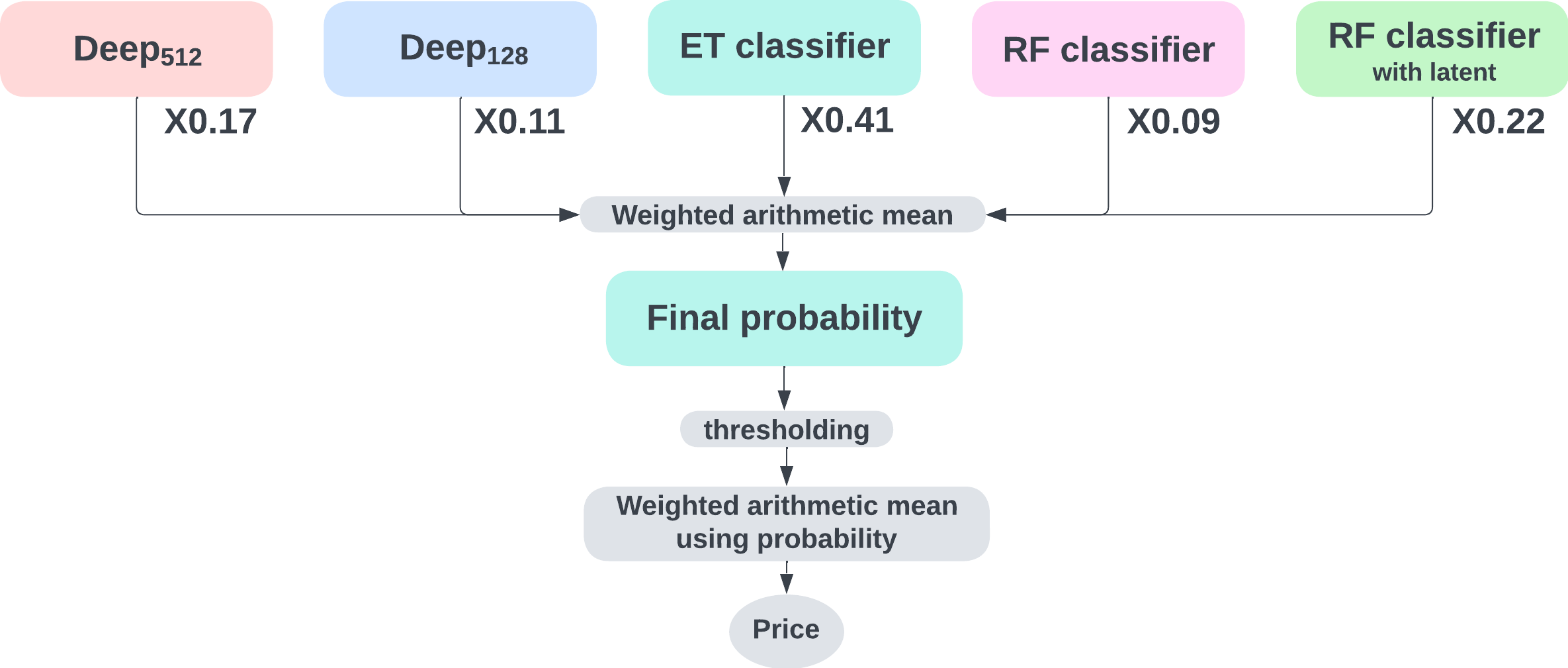}}
\caption{Illustration of the ensemble technique and the process of calculating the predicted price.}
\label{fig4}
\end{figure}

\indent
To combine the similarity scores from these five models, we used a weighted averaging approach. The weights for each model were fine-tuned using Optuna, which allowed us to optimize the ensemble model's performance. The Optuna optimization process involved varying the weights of each model and finding the combination that resulted in the best performance from the validation set. The weight for each model is represented in Fig.~\ref{fig4}.

\section{Results and Discussion}
The evaluation of the results for the test dataset was conducted using two metrics, namely the Area Under the ROC Curve (AUC) for the classification part, and the Mean Absolute Percentage Error (MAPE) for evaluating the final predicted price. AUC was selected as the performance metric due to its ability to evaluate the model's performance across various classification thresholds, making it suitable for real-world applications. Additionally, MAPE measures the accuracy of the price predictions in terms of percentage error, providing insight into how much the model's predicted prices differed from the actual prices on average. Using both of these metrics created a strong evaluation method that supports the usability of the model.

\indent
The baseline model, Extra Trees Classifier, which was trained without image features, achieves an AUC of approximately 0.74. The best model after ensembling outperforms our baseline model, achieving an AUC of approximately 0.81. We report detailed results for other models in Table~\ref{table1}. In the ensemble model, the hyperparameters of the deep learning model from the optimization process are reported in Table~\ref{table2}.

\begin{table}[htbp]
    \caption{The AUC results of each model on test set}
    \begin{center}
        \begin{tabular}{|l|c|} 
        \hline
        \multicolumn{1}{|c|}{\textbf{Models}} & \textbf{Test AUC} \\
        \hline
        Extra Trees Classifier without image features \textbf{(baseline)} & 0.739 \\
        \hline
        Extra Trees Classifier with image features & 0.753 \\
        \hline
        EfficientNet-b1 & 0.780 \\
        \hline
        Random Forest Classifier (trained with latent space) & 0.789 \\
        \hline
        Ensemble model & \textbf{0.812} \\
        \hline
    \end{tabular}
    \end{center}
    \label{table1}
\end{table}

\begin{table}[htbp]
    \caption{The description of hyperparameters and settings of the deep learning model}
    \begin{center}
        \begin{tabular}{|l|l|}
        \hline
        \multicolumn{1}{|c|}{\textbf{Hyperparameter/Setting}} & \multicolumn{1}{|c|}{\textbf{Value}} \\
        \hline
        Model Architecture & EfficientNet-b1 \\
        \hline
        Partial Layer Freezing & First 7 MBConv blocks \\
        \hline
        Transfer Learning & ImageNet \cite{ImageNet} \\
        \hline
        Loss Function & Binary Cross Entropy \\
        \hline
        Optimizer & SGD with Nesterov Momentum = 0.9 \\
        \hline
        Initial Learning Rate & 0.025 \\
        \hline
        Learning Rate Scheduler & Cosine Annealing Warm Restarts \cite{scheduler} \\
        \hline
        Dropout & 0.07 \\
        \hline
        No. of Hidden Layers Nodes & 400, 200, 100, 50 \\
        \hline
        Image Size & 512 \\
        \hline
        Batch Size & 64 \\
        \hline
    \end{tabular}
    \end{center}
    \label{table2}
\end{table}

\begin{figure}[htbp]
\centerline{\includegraphics[width=1.0\columnwidth]{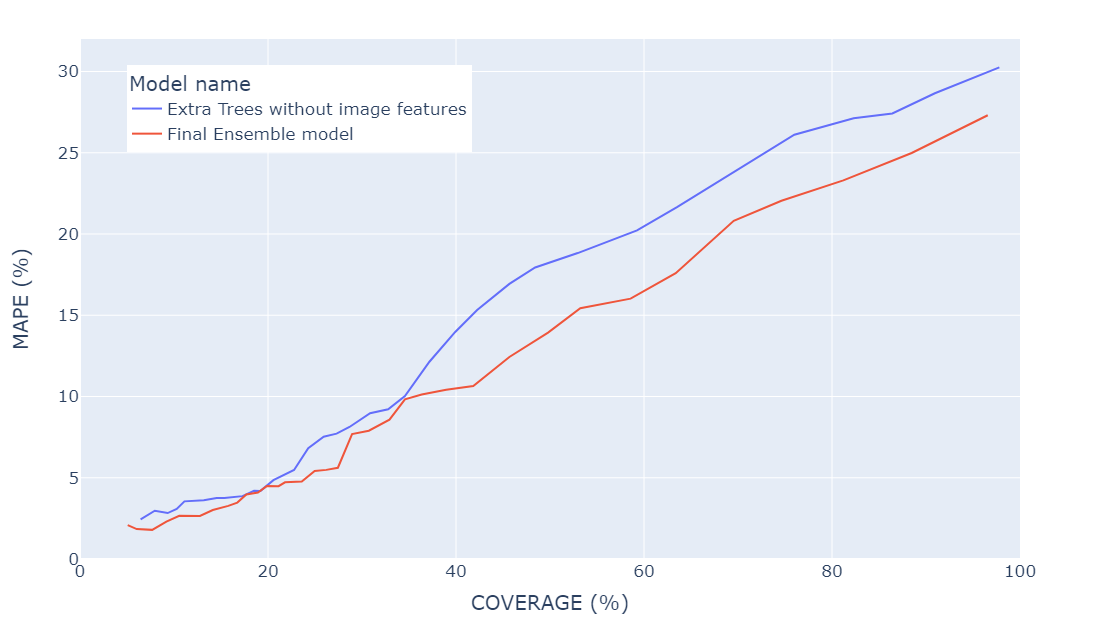}}
\caption{The comparison of Coverage-MAPE performance between the Final Ensemble model and the Extra Trees Classifier trained without image features (baseline model).}
\label{fig5}
\end{figure}

\indent
As shown in Fig.~\ref{fig5}, for the evaluation of MAPE, we compared the models by fixing the MAPE at 20\% and determined the Coverage Percentage, which has a similar interpretation to recall. Recall, also known as sensitivity or true positive rate, is a measure of the proportion of true positive cases out of the total actual positive cases. In our case, the Coverage Percentage represents the proportion of lands that had at least one nearby land predicted as similar by the model, divided by the total number of cases with positive predictions.

\indent
We found that our final best model is able to improve the Coverage Percentage from 59.26\% (baseline model) to 69.55\%, which is a significant improvement. This indicates that our model is able to identify more similar nearby lands and provide predicted prices for them, compared to the baseline model. It should be noted that setting a minimum threshold for the similarity score would result in some lands not being predicted as positive cases and thus not being assigned a predicted price. Therefore, the higher Coverage Percentage achieved by our best model implies that more lands were accurately predicted and, hence, assigned reasonable prices.

\indent
Overall, the results indicate that our ensemble model, which utilizes aerial or satellite imagery, outperformed the baseline model in terms of both classification performance (AUC) and predicted price accuracy (MAPE). This suggests that our approach of leveraging the strengths of multiple models through similarity learning and ensemble techniques has resulted in a more accurate and reliable asset valuation model for predicting land prices. 

\indent
For comparison, we also created a regression model using Gradient Boosting Tree \cite{gbr}, which performed the best in a straightforward regression setting, achieving an MAPE of 34.15\%, notably higher than our final model results. This highlights the importance of changing the way we approach the problem, from a traditional regression-based approach to a classification-based approach, to develop an accurate land price prediction model.

\begin{figure}[htbp]
\centerline{\includegraphics[width=1.0\columnwidth]{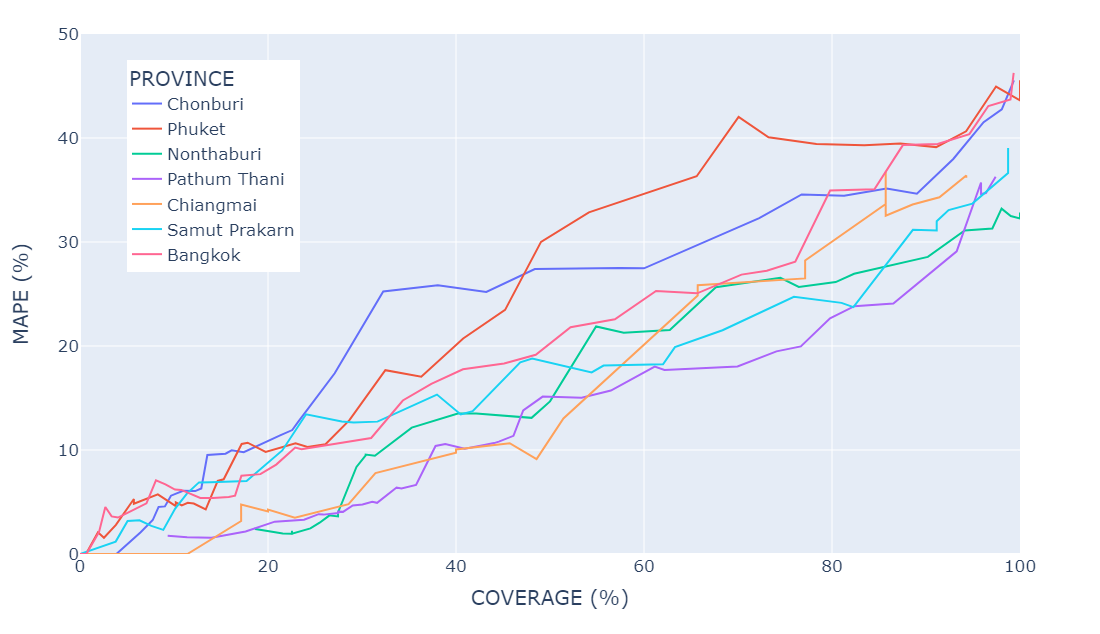}}
\caption{The comparison on Coverage-MAPE performance of each province.}
\label{fig6}
\end{figure}

\indent
For more detailed analysis, from Fig.~\ref{fig6}, we observe that the model performs well for most of the provinces, with low MAPE at each Coverage Percentage. However, two provinces, Phuket and Chonburi (South and Central of Thailand respectively), in particular, show substantially worse results compared to the others. Further investigation into the causes of the poor performance in Phuket and Chonburi revealed that these provinces have considerably less data compared to the provinces around Bangkok. This data scarcity could have negatively affected the model's ability to accurately predict the land prices in these provinces. Additionally, other factors such as the difference in the price of land in Thailand, especially in the tourist-friendly province, and industrial areas, could have also contributed to the poorer performance in these areas.

\indent
The results of the error analysis suggest that data scarcity and the variation in land prices across different areas of Thailand can negatively impact the model's performance. Therefore, future work could focus on obtaining more data from these provinces and exploring other factors that could affect land prices to improve the model's accuracy in these areas.

\section{Conclusion}
In this study, we developed a novel asset valuation model for predicting land prices using similarity learning and deep learning techniques, specifically combining both deep learning and tree-based models through ensemble techniques. By leveraging both tabular data and satellite imagery, we achieve higher accuracy and more reliable predictions than the straightforward end-to-end regression model.

\indent
Our study's findings provide evidence that incorporating satellite imagery and our innovative approach, which utilizes similarity scores instead of directly predicting prices from images and tables, can noticeably enhance the effectiveness of land price prediction. By evaluating the model on a test dataset, we observed that our best model, which combined the similarity scores obtained from deep learning models and tree-based models, outperformed the baseline model in terms of both classification performance (AUC) and predicted price accuracy (MAPE). The AUC of our best model was approximately 0.81, indicating a higher level of classification performance compared to the baseline model. Moreover, the model was able to improve the Coverage Percentage from 59.26\% (baseline model) to 69.55\%, indicating a higher accuracy in predicting the reasonable prices of land.

\indent
It is important to acknowledge the limitations of our study. The primary limitation is the scarcity of data. Since our approach heavily depends on utilizing the similarity score of the neighboring regions, the absence of data about the surrounding areas can adversely affect the accuracy of our land price prediction. Hence, further research studies that incorporate larger and more diverse datasets are necessary to validate the efficacy of our model. Furthermore, our research was limited to the geographical region of Thailand, which enhances the effectiveness of our method as it aligns with the underlying principles of the dataset.

\section*{Acknowledgment}

We would like to express our gratitude to Kasikorn Business Technology Group (KBTG) for providing us with the valuable land data used in our study, as well as for the computing resources for deep learning training. Without their support, this research would not have been possible. We would also like to extend our thanks to Mr. Techin Techintananan, who helped us in scraping the land boundary data that was essential for our data preprocessing. Finally, we would like to thank all of our colleagues who have provided us with valuable support throughout the project.

\bibliographystyle{IEEEtran}
\bibliography{ref} 

\end{document}